\documentclass[pre,11pt]{revtex4-1}
\usepackage{amssymb,amscd,amsmath}
\usepackage{graphicx}
\usepackage{xcolor}
\usepackage{mathtools}
\usepackage{float}
\usepackage{kotex}
\usepackage{lineno}
\def\dm{\Delta_\text{m}}
\def\pps{p'_s}
\def\ncp{N_\text{cp}}
\def\pms{P_m^*}
\def\pml{L_m^*}
\def\pss{p_s^*}
\def\psav{\overline{p_s}}
\def\psavg{\ave{\psav}_g}
\def\pds{p_d^*}
\def\psa{p_s^A}
\def\psb{p_s^B}

\def\xbg{X_g^B}

\def\cb{C^B}
\def\cbt{\tilde{C}^B}
\def\cbg{C_g^B}

\def\mpa{\tilde{M}^{A}}
\def\fpa{\tilde{F}^{A}}
\def\mpb{\tilde{M}^{B}}
\def\fpb{\tilde{F}^{B}}
\def\beqa{\begin{eqnarray}}
\def\eeqa{\end{eqnarray}}
\def\half{\frac{1}{2}}
\def\a={&=&}
\def\b={\ =\ }
\def\sa={\hsg &=& \hsg}
\def\tm{truemm}
\def\hsg{\mbox{} \hsp{-2} }
\def\hsp#1{\hspace{#1 \tm}}
\def\vsp#1{\vspace{#1 \tm}}
\def\aad{A_\text{ad}}
\newcommand{\pd}[2]{\frac{\partial {#1}}{\partial {#2}}}
\newcommand{\ave}[1]{\langle {#1} \rangle}
\newcommand{\Ave}[1]{\left\langle {#1} \right\rangle}
\newcommand{\pa}[1]{\left( {#1} \right)}
\newcommand{\pas}[1]{\left[ {#1} \right]}

\begin{document}
\title[]{Balancing the Last Birth: A Game-Theoretical Resolution to the Human Sex Ratio Puzzle}
\author{Nahyeon Lee}
\affiliation{Department of Physics, Sejong University, Seoul 05006, Korea}
\author{Ulf Dieckmann}
\affiliation{Complexity Science and Evolution Unit,
  Okinawa Institute of Science and Technology Graduate University (OIST),
  Tancha, Onna, Kunigami, Okinawa 904-0495, Japan
  }
\author{Hyeong-Chai Jeong}
\email[]{hcj@sejong.ac.kr}
\affiliation{Department of Physics, Sejong University, Seoul 05006, Korea}
\affiliation{School of Computational Sciences, Korea Institute for Advanced Study,
  Seoul 02455, Korea}

\date{\today}

\begin{abstract}
We study the evolution of offspring sex ratios using a game-theoretical model
in which the decision to have another child depends on the sex of the previous child.
Motivated by higher male infant mortality and the tendency to try again after a child's death,
our model allows different continuation probabilities after sons ($b_s$) and daughters ($b_d$).
We find that a stable sex ratio at birth (SRB) differing from 1:1 can arise when $b_s \neq b_d$.
However, the sex ratio among last-born children (SRLB) always converges to 1:1.
We mathematically prove that this 1:1 SRLB is an evolutionarily stable strategy under
a new fitness measure based on the number of offspring in successful mating pairs,
rather than the number of descendants in the whole population.
Our results generalize Fisher’s principle by showing that equilibrium is maintained
at the level of last births even when the overall SRB is biased.
This offers a potential explanation for the persistent slight male bias in human births,
linking it to sex-specific child mortality and
parental reproductive strategies in historical populations.
\end{abstract}

\maketitle

\section{Introduction}
The evolution of sex ratios has long been a fundamental topic in evolutionary biology.
One central question is why most animals produce roughly equal numbers of sons and daughters,
a question first raised by Darwin and later formalized by Fisher as
Fisher’s principle \cite{fisher1930genetical}.
According to this principle, if one sex becomes less common in the population,
individuals of that sex gain a reproductive advantage.
As a result, genes favoring the production of the rarer sex will spread,
eventually restoring a balanced sex ratio.
Under assumptions such as random mating and equal cost of producing sons and daughters,
this leads to a stable 1:1 sex ratio.

However, the actual sex ratio at birth (SRB) in humans is not exactly equal.
It is consistently male-biased, with about 105 boys born for every 100 girls.
This pattern has remained stable across time and
cultures\cite{chahnaz2021historical, guilmoto2012sex, bongaarts2013trends}.
This raises the question of whether this deviation is within the expected range of natural
variation or whether it reflects a deeper evolutionary or behavioral mechanism.

Several evolutionary theories have been proposed to explain deviations from the 1:1 sex ratio.
These include local resource competition (LRC)\citep{clark1978sex},
local mate competition (LMC)\citep{hamilton1967extraordinary},
local resource enhancement (LRE)\citep{gowaty1993local},
and condition-dependent sex allocation\citep{trivers1973natural}.
These theories suggest that sex ratios can shift depending on ecological and social factors.

However, most of these models tend to predict female-biased sex ratios under typical conditions,
which is opposite to the human case.
In addition, applying these theories directly to humans is not always straightforward,
as human societies have complex cultural, medical, and behavioral factors that also
influence reproduction.

In this paper, we revisit the question of sex ratio evolution using a game-theoretical model.
We examine how differences in the likelihood of having another child after a son versus a daughter
can affect the long-term stable sex ratio.

We propose that this difference may arise from sex-specific child mortality,
as boys generally face a higher risk of dying than girls.
If parents are more likely to have another child when a previous child has died,
then this mortality difference could lead to a higher chance of having another child
after a son than after a daughter.
This behavior can be captured in our model by assigning different continuation probabilities,
represented by the birth probabilities $b_s$ and $b_d$, after sons and daughters, respectively.

We study the evolution of a genetic trait called the ``son probability",
denoted by $p_s$, which represents the probability of producing a male child.
We find that this trait evolves to a stable value given by
$p_s^* = \frac{1 - b_d}{2 - b_s - b_d}$.
This value is not necessarily $1/2$,
meaning that the sex ratio at birth (SRB) can be biased unless $b_s = b_d$.
Interestingly, even when the SRB is not 1:1,
this evolved value of $\pss$ leads to a balanced sex ratio among the last-born children (SRLB).

We prove that an SRLB of 1:1 is an evolutionarily stable strategy (ESS)
when fitness is defined as the expected number of offspring who eventually form couples.
In contrast, if fitness is defined as the expected number of descendants in the whole population,
whether we use children, grandchildren, or great-grandchildren,
we cannot mathematically prove the stability of the strategy leading to SRLB.
However, our numerical simulations show that individuals with the gene value
$p_s^*$ tend to have the largest number of descendants in the long run,
supporting the validity of defining fitness as the expected number of offspring who eventually form couples.

The rest of this paper is structured as follows.
In Section II, we define the model and key parameters,
focusing on the dynamics and inheritance mechanisms.
Section III derives the steady-state sex ratio, showing that it satisfies 1:1 SRLB.
In Section IV, we analyze evolutionary stability, proving that SRLB is an ESS under
gene-based fitness evaluation.
In Section V we apply our observation to the human sex ratio puzzle. 
Finally, Section VI summarizes our findings and suggests directions for future research.

\section{Model and Methods}
We start with a population consisting of an equal number of males and females,
denoted by $N$ for each.
Each individual has a ‘son probability’, $p_s$,
which represents the probability of producing a male offspring (son).
We study the evolution of this son probability,
along with inheritance and mutation dynamics.
In our model, the probabilities of having another child after a male or female offspring
may differ and are denoted by $b_s$ and $b_d$, respectively.
These probabilities remain constant throughout the evolution of the sex ratio.

Males and females are randomly paired to form couples.
Each couple produces at least one offspring,
with the gender determined by the father's $p_s$.
The probability of producing a male offspring is $p_s$,
while the probability of producing a female offspring is $1 - p_s$.
The $p_s$ value of the offspring is inherited from either parent with equal probability.
Mutations occur at a small rate, $\mu$,
introducing slight variations of $\pm \delta$
under the constrant that $p_s$ stays within the range of 0 and 1.

The probability of producing additional offspring depends on the gender of
the previous child, $b_s$ if the previous child was male, and $b_d$ if female.
This process continues until the couple stops reproducing.

A generation refers to one complete cycle of these population processes.
During each generation, $N$ pairs of males and females are randomly matched,
and each couple produces offspring according to the probabilities $b_s$ and $b_d$.
The cycle continues until all pairs have completed reproduction.
The process repeats across successive generations to observe changes
in population characteristics over time
until the population reaches a steady state where
population properties remains almost constant
in the range of equilibrium fluctuations.

\section{Evolution of Sex Ratio}

We begin with a population of 200 adults, consisting of $N_M = 100$ males and $N_F = 100$ females
at the initial generation ($g = 0$).
Each individual's son probability, $p_s$, is drawn from a uniform distribution between 0 and 1 at $g = 0$.

\begin{figure}
\includegraphics[height=37.5\tm,width=60\tm]{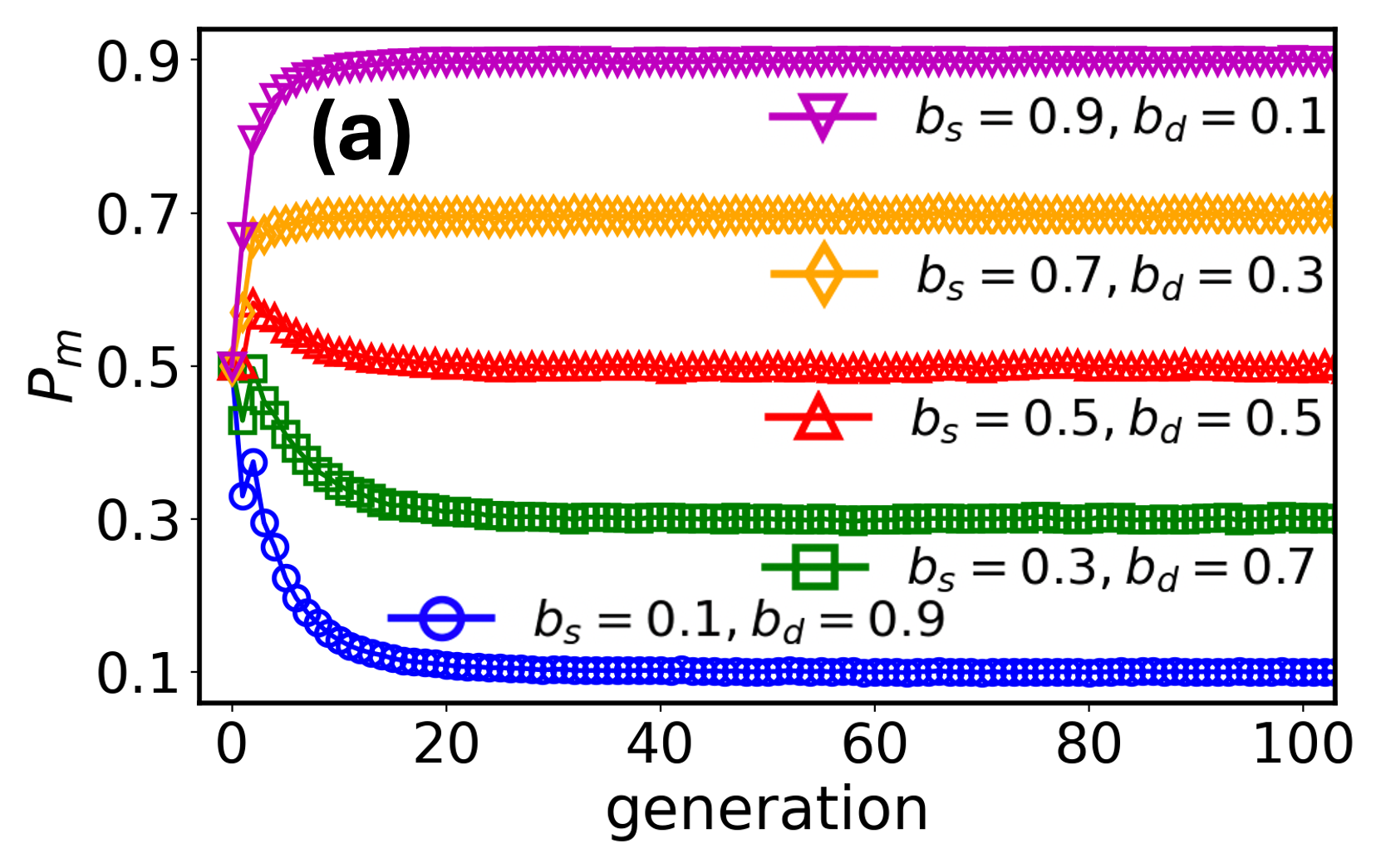}\\
\includegraphics[height=37.5\tm,width=60\tm]{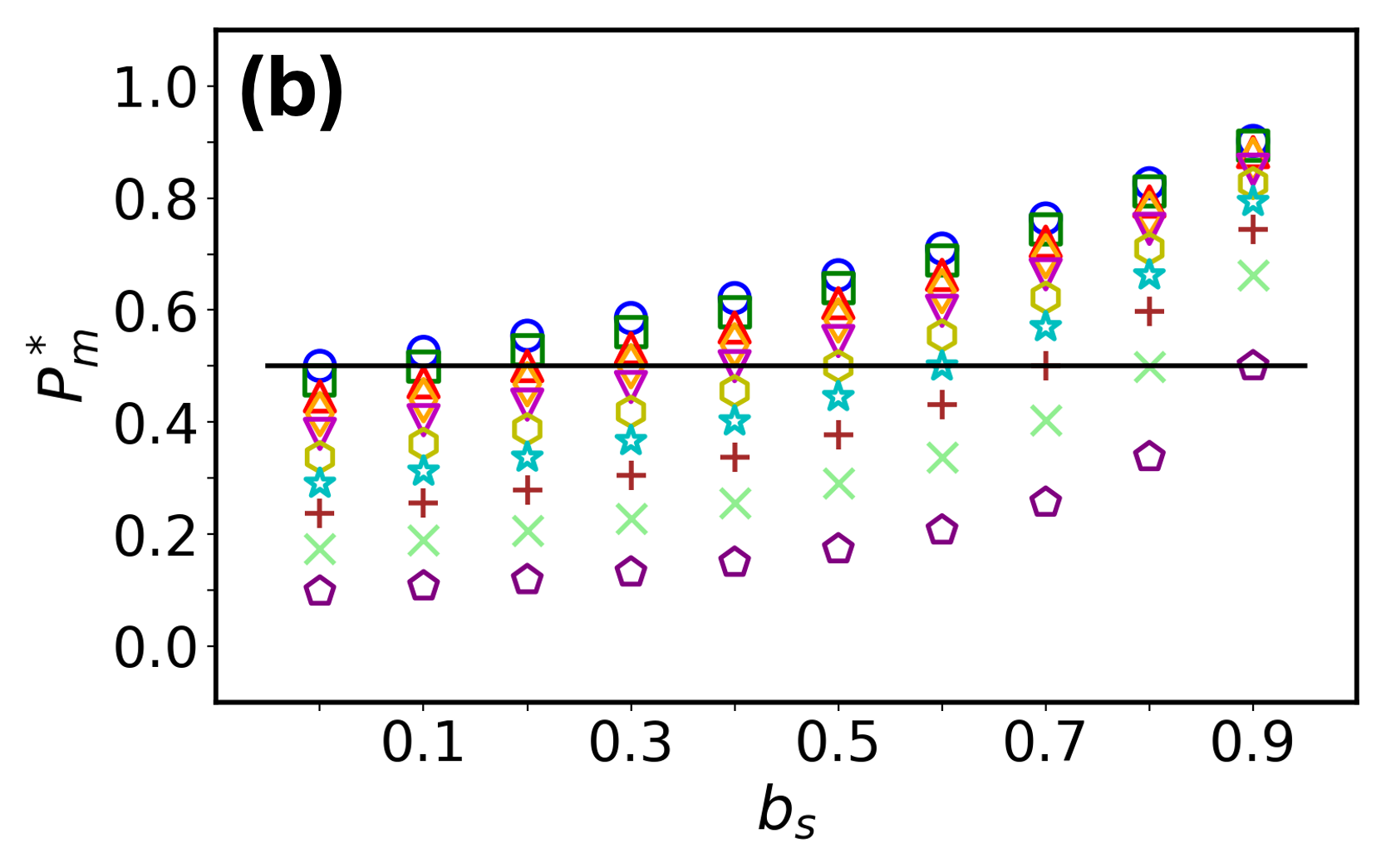}\\
\includegraphics[height=37.5\tm,width=60\tm]{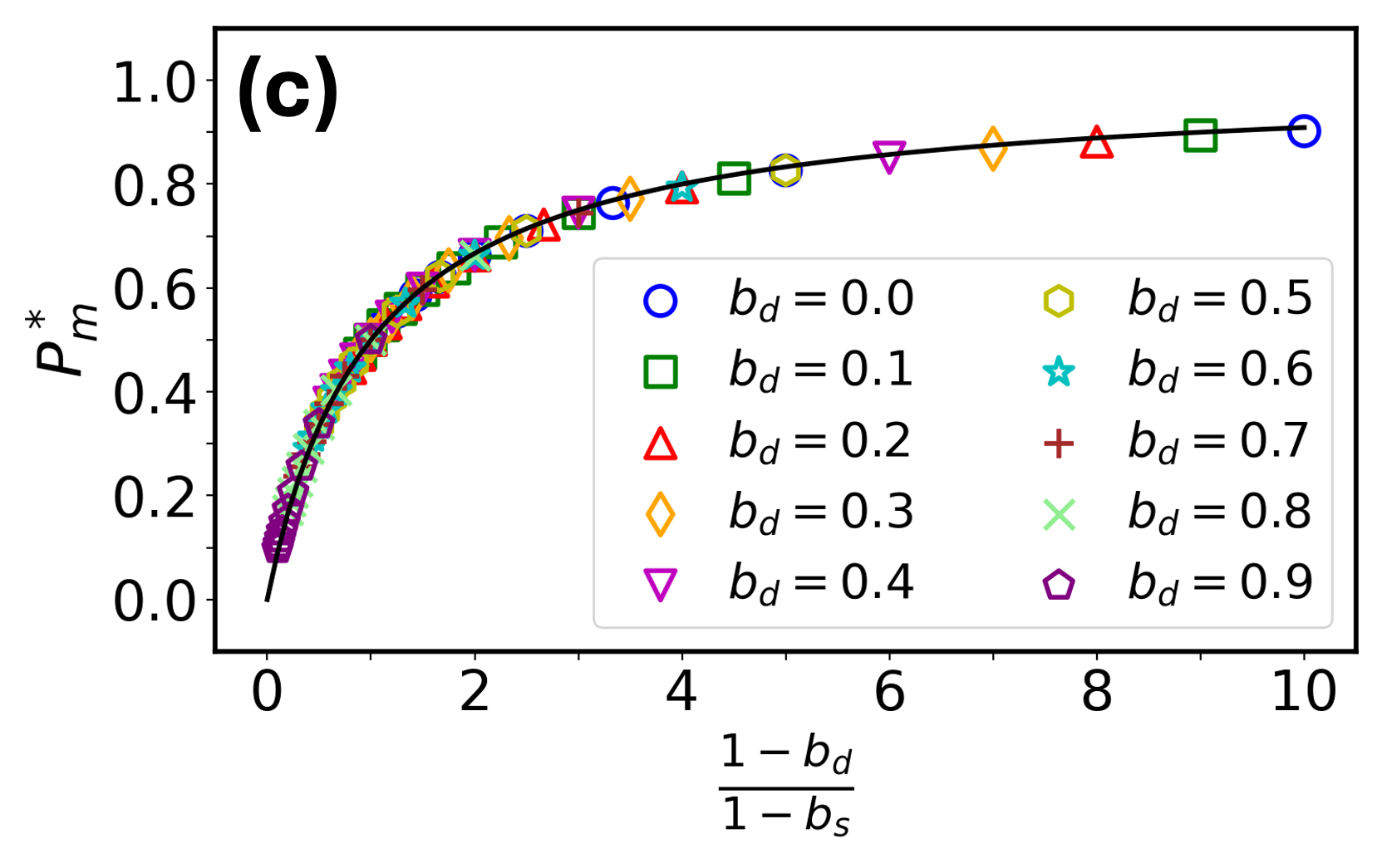}\\

\caption{
  (a) Male ratio, $P_m$, as a function of generation, $g$.
  (b) Male ratio at steady state, $\pms$, as a function of $b_s$ for 10 different values of $b_d$.
  (c) $\pms$ as a function of $\frac{1-b_d}{1-b_s}$ for the same 10 values of $b_d$.
  Each dataset represents the average of $M=10^3$ independent simulations
  with $\ncp=100$ couples.
  Figures (b) and (c) share the legend in (c).
  }
\label{f-pm}
\end{figure}

Figure~\ref{f-pm}(a) presents the male ratio, $P_m$, defined as
\beqa
P_m
\a= \Ave{\frac{N_M}{N_M + N_F}}
\eeqa
as a function of generation $g$ for five different sets of $b_s$ and $b_d$,
$(b_s, b_d) = (0.9, 0.1)$, (0.7, 0.3), (0.5, 0.5), (0.3, 0.7), and (0.1, 0.9)
where $N_M$ and $N_F$ are the number of males and females in the population.
The results are averaged over $M = 10^3$ independent simulations.
Initially, $P_m$ is 0.5 in all cases since $N_M = N_F$ at $g = 0$.
As $g$ increases, $P_m$ rises when $b_s>b_d$ and decreases when $b_s<b_d$.
Over time, $P_m$ converges to steady-state values,
which differ depending on the specific values of $b_s$ and $b_d$.
 
Figure~\ref{f-pm}(b) shows the male ratio at steady state, $\pms$,
averaged over $5 \times 10^5$ generations in the steady state.
The values are computed for 100 different combinations of $b_s$ and $b_d$.
The data is plotted as a function of $b_s$ for ten fixed values of $b_d$,
ranging from 0.0 to 0.9 in increments of 0.1.

For a given $b_d$, $\pms$ increases as $b_s$ grows,
while for a given $b_s$, it decreases as $b_d$ grows.
When $b_s = b_d$, where the probability of the next birth is independent of
the sex of the previous offspring, $\pms=1/2$, consistent with Fisher's argument.
Interestingly, these ten curves collapse into a single curve when $\pms$ is
plotted against $x = (1-b_d)/(1-b_s)$, as shown in Fig.~\ref{f-pm}(c).
The resulting curve is well-fitted by the function $x/(1 + x)$,
represented by the solid line in the figure.
Thus, the male ratio at steady state can be expressed as
\beqa
\pms \a= \frac{1-b_d}{2-b_s-b_d}.
\label{e-pms}
\eeqa

Since $P_m$ is determined by $p_s$, we believe that
the average son probability in the steady state,
$\pss$, should coincide with $\pms$, i.e.,
\beqa
\pss \a= \frac{1-b_d}{2-b_s-b_d}.
\label{e-pss}
\eeqa

In fact, we measure the evolution of $p_s$ distribution
and confirm that $\pss$ and $\pms$ agree within statistical error
as presented in Appexdix~A. 
Furthermore, in the limit of a large population,
we believe that the $\pss$ distribution to form a narrow peak centered at
$\pms$, with the width approaching zero as the mutation rate decreases to zero.
Similarly, if the population starts with a uniform $p_s$ distribution
and evolves without mutation, we expect the population to converge to
$\pms$ in the large population limit.
For finite populations, stochastic effects may cause fixation at values
deviating from $\pms$, but the average fixation value would still
align with $\pms$ as discussed in Appendix~A.
Later, we will demonstrate that $\pss$ in Eq.~(\ref{e-pss}) is
an Evolutionarily Stable Strategy (ESS).
Before that, let us first explore its implications.

Equation~(\ref{e-pss}) can be rewritten as
\beqa
 \pss\pa{1-b_s}
 \a= \pds\pa{1-b_d},
\label{e-srlb} 
\eeqa
where $\pds = 1 - \pss$ represents the probability that an offspring is a daughter.
On the left-hand side, $\pss$ is the probability of having a son,
and $(1-b_s)$ is the probability of not having another child.
Together, the left-hand side represents the probability that
the last child is a son.
Similarly, the right-hand side represents the probability that
the last child is a daughter.
This implies that, in the steady state, the sex ratio at last birth (SRLB)
should be 1:1.

When $b_s=b_d$, the terms $(1-b_s)$ and $(1-b_d)$ cancel out,
reducing Eq.~(\ref{e-srlb}) to $\pss = \pds$, or $\pss = 1/2$.
This corresponds to a 1:1 sex ratio at birth (SRB),
consistent with Fisher's principle~\cite{fisher1930genetical}.
However, if $b_s\neq b_d$, Eq.~(\ref{e-srlb}) indicates that it is the SRLB,
rather than the SRB, that should be 1:1 in the steady state.

\begin{figure} \vsp{2}
  \includegraphics[height=37.5\tm,width=60\tm]{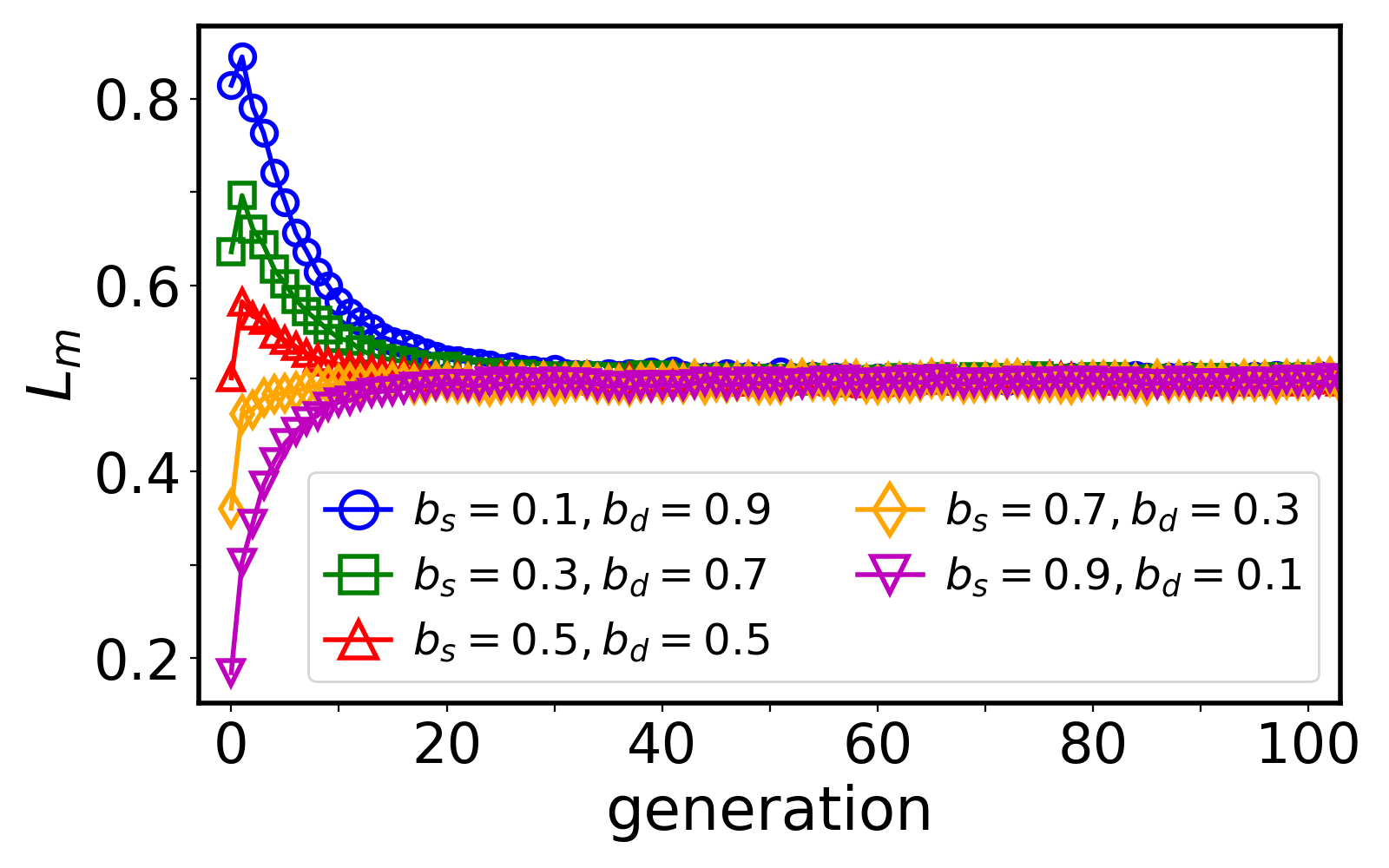}\\
  \caption{
    Evolution of the Male Ratio at Last Birth, $\pml$, Over Generations.
    The average male ratio at last birth, $L_m$, is shown for five different
    $(b_s, b_d)$ pairs, (0.1, 0.9), (0.3, 0.7), (0.5, 0.5), (0.7, 0.3), and (0.9, 0.1).
    After approximately 60 generations, all cases reach a steady-state value of 0.5.
    Each dataset represents the average of $M = 10^3$ independent simulations with $\ncp = 100$ couples.
  }
  \label{f-srlb}
\end{figure}

Figure~\ref{f-srlb} illustrates the evolution of $L_m$,
the average male ratio at the last child, for five different sets of
$b_s$ and $b_d$ values,
$(b_s, b_d) = (0.1, 0.9)$,
(0.3, 0.7), (0.5, 0.5), (0.7, 0.3), and (0.9, 0.1).
In all cases, $L_m$ converges to a steady-state value, $\pml$, of 0.5.

\section{Evolutionary Stability of an Even SRLB}
We now analyze the evolutionary stability of the population consists with
$\pss(b_s,b_d)$ of Eq.~(\ref{e-pss}). 
To do this, we consider a population initially consisting entirely of
$\pss$ and introduce an invading strategy, $\pps = \pss + \Delta$.
For the evolutionary stability analysis,
we compare the relative fitness of $\pps$ to that of $\pss$.
Fitness is typically defined as the expected number of offspring produced by an
individual.
However, for sex-related traits, such as genes influencing the probability of
producing male offspring, defining fitness as the expected number of offspring
is insufficient.
Reproductive success depends not only on offspring quantity but also on their sex
and the population's sex ratio.
To address this, fitness can be more accurately defined as the expected number of
grandchildren, reflecting the long-term reproductive contribution of an individual's
offspring~\cite{fisher1930genetical,frank1990sex,argasinski2008problems,Gardner2019}.

For the expected number of grandchildren, or more broadly, for the expected number
of descendants over subsequent generations with the two traits
$\psa = \pss$ and $\psb = \pps$, we use recursion equations.
These equations describe the evolution of population composition across generations
by updating the counts of traits based on reproduction and inheritance mechanisms.
We express $\mpa$ and $\fpa$, the expected numbers of males and females with trait A
in generation $g+1$, and $\mpb$ and $\fpb$, the expected numbers of males and
females with trait B in generation $g+1$,
in terms of the population compositions in generation $g$ as follows.
\beqa
 \frac{\mpa}{\ncp}
  \sa= m^A f^A n^A_\text{ch} p_s^A
     + \half\pa{m^A f^B n^A_\text{ch} p_s^A  + m^B f^A n^B_\text{ch} p_s^B} 
     \nonumber \\
 \frac{\fpa}{\ncp}
  \sa=  m^A f^A n^A_\text{ch} p_d^A
     +  \half\pa{m^A f^B n^A_\text{ch} p_d^A  + m^B f^A n^B_\text{ch} p_d^B}
     \nonumber \\
 \frac{\mpb}{\ncp}
  \sa= m^B f^B n^B_\text{ch} p_s^B
     + \half\pa{m^B f^A n^B_\text{ch} p_s^B  + m^A f^B n^A_\text{ch} p_s^A}
     \nonumber \\
 \frac{\fpb}{\ncp}
  \sa= m^B f^B n^B_\text{ch} p_d^B
     + \half\pa{m^B f^A n^B_\text{ch} p_d^B  + m^A f^B n^A_\text{ch} p_d^A }
     \nonumber \\
\label{e-gevo}     
\eeqa
Here, 
$p_d^{A,B} = 1 - p_s^{A,B}$ denotes the probability that an offspring is a daughter
when the father has trait A ($p_d^A$) or B ($p_d^B$),
and $n^{A,B}_\text{ch} = \frac{1}{1 - \pa{p_s^{A,B} b_s + p_d^{A,B} b_d}}$ represents
the expected number of offspring produced by a couple
where the husband possesses trait A or B.
The terms $m^{A,B} = \frac{M^{A,B}}{M^A+M^B}$ and $f^{A,B} = \frac{F^{A,B}}{F^A+F^B}$ denote
the proportions of males and females with trait A or B in the population
at generation $g$, respectively.

\begin{figure} \vsp{2}
 \includegraphics[height=37.5\tm,width=60\tm]{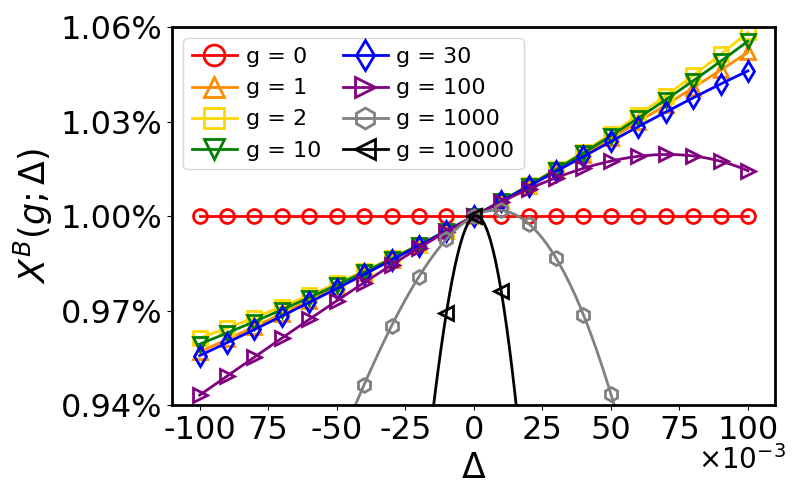}\\
 \caption{
   Ratio of $B$ genes with $\psb(\Delta) = \pss + \Delta$
   in the whole population, $\xbg$,
   as a function of $\Delta$ at eight different generations,
   $g = 0$, 1, 2, 10, 30, $10^2$, $10^3$, and $10^4$, with $b_s=0.7$ and $b_d=0.3$.
   At $g = 0$, 1\% of the population carries gene B in both male and female groups.
 }
\label{f-xb}
\end{figure}

Figure~\ref{f-xb} shows the ratio,
\beqa
 \xbg
 \a= \frac{N^B_g}{N^A_g+N^B_g}
\label{e-xb} 
\eeqa
of B genes with $\psb(\Delta)=\pss+\Delta$
in the population of generation~$g$
as a function of $\Delta$, where $N^{A,B} = M^{A,B} + F^{A,B}$.
Initially the population consists of 99 percent of $\psa=\pss$
and 1 percent $\psb$ and evolves with Eq.~(\ref{e-gevo}).

We observe that $\xbg$ increases monotonically with $\Delta$ within the region shown in the figure,
not only in the first generation ($g=1$) but also in the second generation ($g=2$).
This indicates that $\pss$ does not qualify as an ESS in the traditional sense,
which requires the frequency of any invading gene to decrease by the second generation (grandchild generation)
when introduced into a population of $\pss$.
However, over subsequent generations, $\xbg$ for values other than $\Delta = 0$ gradually decreases
and eventually falls below its value at $\Delta = 0$.
For example, as shown in Fig.~\ref{f-xb}, at $g = 10^4$, $\xbg$ reaches its maximum at $\Delta = 0$.
Therefore, if fitness is defined as the gene frequency in the population at sufficiently later generations
(instead of at the second generation), it might be possible to show that $\pss$ is an ESS.
However, this definition introduces arbitrariness regarding the choice of the specific generation used.

If we define fitness based on the ratio of genes in matched pairs rather than
in the entire population, we find that $\pss$ is an ESS.
This means that no individual carrying a gene with a different son probability
than $\pss$ can achieve a higher fitness when introduced into
a population where all individuals have the gene $\pss$.
We again consider a population that initially consists of 99\% individuals with
gene $\psa = \pss$ and 1\% individuals with gene $\psb(\Delta) = \pss + \Delta$.
The population evolves according to Eq.~(\ref{e-gevo}), but instead of measuring
the gene ratio in the entire population, we now measure the ratio of $B$ genes
in the matched pairs, denoted as $\cb$.
This is given by,
\beqa
 \cb
 \a= \frac{1}{2}\pa{m^B + f^B} \nonumber \\
 \a= \frac{1}{2}\pa{\frac{M^B}{M^A+M^B} + \frac{F^B}{F^A+F^B}},
 \label{e-cb}
\eeqa
where the factor $\frac{1}{2}$ accounts for the fact that the number of males
and females in matched pairs is equal.

\begin{figure} \vsp{2}
 \includegraphics[height=37.5\tm,width=60\tm]{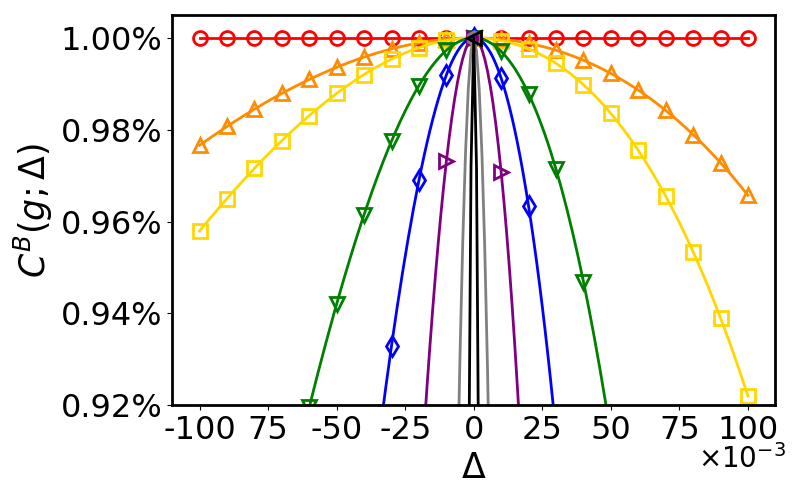}\\
 \caption{
   The ratio of $B$ genes with $\psb(\Delta) = \pss + \Delta$
   in the matched pairs, $\cbg$, is shown as a function of $\Delta$
   at eight different generations,
   $g = 0$, 1, 2, 10, 30, $10^2$, $10^3$, and $10^4$, with $b_s = 0.7$ and $b_d = 0.3$.
   At $g = 0$, 1\% of the population carries gene B in both male and female groups.
   This figure shares the same legend as Figure~\ref{f-xb}.
 }
 \label{f-cb}
\end{figure}

In Fig.~\ref{f-cb}, we show $\cbg$ as a function of $\Delta$ for various
generations. Here, $m^{B}$ and $f^{B}$ are calculated using Eq.~(\ref{e-gevo}),
starting with the initial values $m^{B}=1/100$ and $f^{B}=1/100$ at $g=0$.
From the figure, we observe that
$C^B(g, \Delta \not= 0)$ is always smaller than $C^B(g, \Delta = 0)$
from the very first generation and becomes even smaller as
generations progress. 

In fact, we can analytically demonstrate that $\pss$ is the only ESS
when fitness is defined as the ratio of genes in the matched pairs,
as outlined in Appendix~B.
Since parental genes determining the offspring's sex are inherited
regardless of the offspring's sex, 
the ratios of $B$ genes in males and females are expected to converge
over generations.
Once the ratios $m^B$ and $f^B$ in males and females become equal,
$C^B$ as defined in Eq.~(\ref{e-cb}) becomes identical to $X^B$ in
Eq.~(\ref{e-xb}).
Consequently, we can conclude that 
$\xbg$ reaches its maximum at $\Delta = 0$ 
at sufficiently later generations.

\section{Application to Human Sex Ratio}
In this section, we examine whether the human sex ratio at birth (SRB) of
105 boys to 100 girls~\cite{orzack2015human,chao2019systematic,cdc2005ustrends}
can arise naturally when parents are more likely to have another child if the previous child has died,
and when boys have a higher mortality than girls.

Historical data suggest that in the distant past, about 50\% of children died before reaching adulthood,
where we define adulthood as surviving to age $\aad = 15$\cite{roser2023mortality}.
It is also well documented that boys had a higher probability of dying before this age than girls,
often by more than 10\%\cite{drevenstedt2008rise,un2024world}.
From these observations, we assume that the probability of surviving to adulthood
is about $S_s = 0.5$ for males and $S_d = 0.55$ for females.

We begin by explaining how the observed values can be related to the parameters $b_s$ and $b_d$ in our model.
We assume the population is in a steady state, meaning that each couple produces, on average,
one daughter who survives to adulthood.
Let $n_\text{d,ad}$ be the average number of adult daughters per couple.
For the population to remain stable, we require $n_\text{d,ad} = 1$.
This condition must hold regardless of the marriage system, since the total number of adult females should
remain constant over time.

Given that the sex ratio at birth (SRB) is 105 boys for every 100 girls, the equilibrium son probability is
\beqa
p_s^*
\a= \frac{1 - b_d}{(1 - b_s) + (1 - b_d)}
\b= \frac{105}{205}
\eeqa
To maintain a stable population, we assume that each couple produces, on average, one daughter who survives to adulthood,
\beqa
n_\text{d,ad}
\a= n_\text{ch} \cdot (1 - p_s^*) \cdot S_d
\b= 1,
\label{e.ndad}
\eeqa
where $S_d = 0.55$ is the survival probability of daughters, and
$n_\text{ch} = \frac{1}{1 - \pa{p_s^* \cdot b_s + (1 - p_s^*) \cdot b_d}}$
is the expected number of children per couple.
Solving these equations yields the values $b_s=0.738$ and $b_d=0.725$.

We propose that the difference in the probability of having another child after a son versus a daughter
may be explained by differences in mortality between boys and girls.
It is reasonable to assume that parents are more likely to have another child
if their previous child has died than if the child is still alive.

Let $b_1$ represent the probability of having another child when the last child is alive,
and $b_2$ when the last child has died.
Let $d_s$ and $d_d$ denote the probabilities that the most recently born child,
if a son or a daughter respectively, dies before the end of the parents’ reproductive period.
Then the effective birth probabilities after a son and a daughter, $b_s$ and $b_d$, are given by
\beqa
b_s \a= (1 - d_s) b_1 + d_s b_2 \nonumber \\
b_d \a= (1 - d_d) b_1 + d_d b_2
\label{e.bsbd}
\eeqa
From these expressions, we obtain the ratio
\beqa
\frac{b_s}{b_d}
\a= \frac{b_1 + d_s (b_2 - b_1)}{b_1 + d_d (b_2 - b_1)}
\b= \frac{1 + d_s \pa{\frac{b_2}{b_1} - 1}}{1 + d_d \pa{\frac{b_2}{b_1} - 1}}
\eeqa
Since $b_2 \ge b_1$, the ratio $b_2 / b_1 \ge 1$.
This implies the following inequality,
\beqa
1 \le \frac{b_s}{b_d} \le \frac{d_s}{d_d}.
\eeqa
In other words, the difference in birth probabilities after sons and daughters is bounded above
by the ratio of their death probabilities.

Using this result, we estimate the maximum and minimum possible values of $\pss$,
based on the constraint in Eq.~(\ref{e.ndad}), which requires that each couple produces,
on average, one daughter who survives to adulthood.
According to this constraint, $\pss$ increases as the ratio $\frac{b_s}{b_d}$ increases.
When $\frac{b_s}{b_d} = 1$, we obtain $\pss = 1/2$, corresponding to a 1:1 SRB.
The maximum possible value of $\pss$ occurs when $\frac{b_s}{b_d} = \frac{d_s}{d_d}$.
Although the exact value of this mortality ratio is uncertain, it is estimated to be about 1.1,
reflecting the observation that boys have roughly 10\% higher mortality than girls before reaching adulthood.

Substituting $\frac{b_s}{b_d} = \frac{d_s}{d_d} = 1.1$ into Eq.~(\ref{e.ndad}) yields $b_s = 0.78$, $b_d = 0.71$,
and $\pss = 0.57$. This corresponds to a sex ratio at birth (SRB) of approximately 133 boys per 100 girls,
which we consider the upper limit predicted by our model.
The lower limit corresponds to $\frac{b_s}{b_d} = 1$, giving the standard 1:1 SRB.

The observed human SRB of 105 boys per 100 girls lies within this theoretical range.
We explain this observation by assuming an intermediate value for the ratio $b_2/b_1$,
which measures how much more likely parents are to have another child if the previous
one has died compared to when the previous child is still alive.
To estimate this ratio, we first consider the case that $d_s = 1 - S_s$ and $d_d = 1 - S_d$,
meaning that if a child dies before reaching adulthood,
it is likely to be the youngest child~\cite{Pozzi2015Infant}.
By substituting $d_s = 0.5$ and $d_d = 0.45$, along with $b_s = 0.738$ and $b_d = 0.725$ for
the observed SRB of 105:100, we calculate $b_1 = 0.59$ and $b_2 = 0.89$.
These results suggest that parents are substantially more likely to have another child
after the death of a previous one, supporting our hypothesis.

In reality, the decision to have another child may depend not only on whether the most recent child is alive or has died, but also on how many children the couple already has.
To account for this, we conducted numerical simulations that track the population year by year.
In these simulations, the probability of having another child decreases as the number of existing children increases.
To model child mortality under a different assumption, we define the annual death probabilities $y_s$ for boys and $y_d$ for girls.
Unlike our earlier analysis, where we assumed that mortality risk mainly affects the youngest child, we now consider the opposite case, every child faces the same constant risk of dying each year until reaching adulthood.
By examining this contrasting assumption, we aim to demonstrate that our explanation for the human sex ratio remains plausible across a reasonable range of assumptions about the distribution of child mortality.
We set the survival probabilities to adulthood as $S_s = 0.5$ for boys and $S_d = 0.55$ for girls
as before. 
Assuming constant annual death probabilities until the age of $\aad = 15$,
which we define as the onset of adulthood,
we use the relations $(1 - y_s)^\aad = S_s$ and $(1 - y_d)^\aad = S_d$
to calculate the annual death probabilities.
This yields $y_s = 0.045$ for boys and $y_d = 0.039$ for girls.

In our simulation, individuals form couples at age 15,
and we assume that each couple consists of one male and one female of the same age.
Every year, new couples are created by pairing all available 15-year-old boys and girls,
with the total number of couples limited by the smaller of the two groups.

Once formed, couples are eligible to have children from age 16 to 40.
The probability that a couple has a child in a given year is modeled as
\beqa
p_b \a= e^{-n} \pa{1 - \frac{N_\text{ch}}{K}},
\label{e-pb}
\eeqa
where $n$ is the number of children the couple already has,
$N_\text{ch}$ is the total number of children under age 15 in the population,
and $K$ represents the population's carrying capacity for children.
This formulation captures the idea that the likelihood of having another child decreases
as a couple has more children and as the total child population approaches the carrying capacity.
We use the same assumptions for sex determination and the inheritance of the son-probability gene
as described earlier in the Methods section for the simpler model with fixed values of $b_s$ and $b_d$.

In the long-term steady state of the simulation,
the average son probability converges to $p_s^* = 0.507$,
which corresponds to a sex ratio at birth (SRB) of about 103 boys for every 100 girls.
Although this is slightly lower than the commonly reported SRB of 105:100,
it demonstrates that a stable SRB different from the classical 1:1 ratio can emerge naturally
when the probabilities of having another child after a son or daughter differ effectively.

We speculate that using more realistic age specific death probabilites,
especially higher mortality during infancy and early childhood,
together with the strong male biased mortality observed in these early
years~\cite{sawyer2012child,aghai2020gender_neonatal},
may produce results that more closely match the known SRB of 105 boys for every 100 girls.

\section{Concluding Remarks}
We have studied a model in which the probability of having another child depends on
the sex of the previous child
and have studied evolution of a genetic trait called of son probability $p_s$,
which represents the probability of producing a male child.

Our assumption that the probability of continuing childbirth depends on the sex of the previous
child is grounded in biological plausibility.
Given that boys generally face higher mortality risks than girls,
and that parents are more likely to have another child following the death of a previous one,
it is reasonable to expect higher continuation after sons than after daughters.
This mechanism is captured in our model through differing birth probabilities after sons ($b_s$)
and daughters ($b_d$), offering a simple but meaningful explanation for asymmetric reproductive behavior.

A major contribution of this study is the generalization of the classical 1:1 prediction
for the sex ratio at birth (SRB). 
Traditional evolutionary theories suggested that a 1:1 SRB is an evolutionarily stable strategy (ESS).
However, we show that a 1:1 sex ratio among last-born children (SRLB) is a more
general and robust equilibrium. 
When $b_s = b_d$, the SRLB and SRB are both 1:1, matching the classical result.
But even when $b_s \ne b_d$, the SRLB remains at 1:1, highlighting its broader applicability.

We also address the issue of how fitness should be defined for sex-related traits.
Previous studies often defined fitness as the gene frequency in the second generation
across the entire population.
However, this raises the question of why the second generation is chosen, rather than the third or later.
To resolve this, we introduce a new fitness based on gene frequency within
reproductively paired individuals.
Since only individuals in couples can produce offspring, this definition is more biologically
relevant and avoids arbitrary dependence on a specific generation.
Under this framework, we could mathematically prove that the gene responsible for a 1:1 sex ratio
among last-born children (SRLB) is an evolutionarily stable strategy (ESS), even when
$b_s$ and $b_d$ differ.

Our theory offers a clear prediction: the sex ratio among last-born children (SRLB)
should be approximately 1:1, even when the overall sex ratio at birth (SRB) deviates from this balance.
This provides a valuable opportunity for empirical validation using population data.
However, in modern human societies, reproductive behavior is influenced not only by biological factors
but also by cultural and social norms.
Therefore, it would be ideal to examine SRLB in historical or pre-modern populations
where these influences were different or less pronounced.

As an alternative, testing this prediction in animal populations may be informative.
In particular, beef cattle under natural breeding conditions may serve as a suitable model.
Cattle are known to have a male-biased SRB of approximately 105 males for every 100 females,
similar to that observed in humans~\cite{laster1973effects,berry2007artificial,seidel2003sexing}.
Investigating the SRLB in such populations could offer insights into whether our
theoretical prediction holds beyond humans.

Finally, while we have treated $b_s$ and $b_d$ as fixed parameters,
future research could explore how these values themselves evolve.
This would allow us to study the coevolution of genetic and cultural factors,
such as sex-determining genes and societal gender preferences,
offering a more complete picture of how sex ratio dynamics develop over time.

\renewcommand{\thefigure}{A\arabic{figure}}
\renewcommand{\theequation}{A\arabic{equation}}
\setcounter{figure}{0}
\setcounter{equation}{0}

\section{Appendix A:  Distribution of Son Probability, $p_s$}

Here, we analyze the evolution of the distribution of
son probability, $p_s$,
that represents the probability of producing male offspring.

\begin{figure}[h!]
 \includegraphics[height=37.5\tm,width=60\tm]{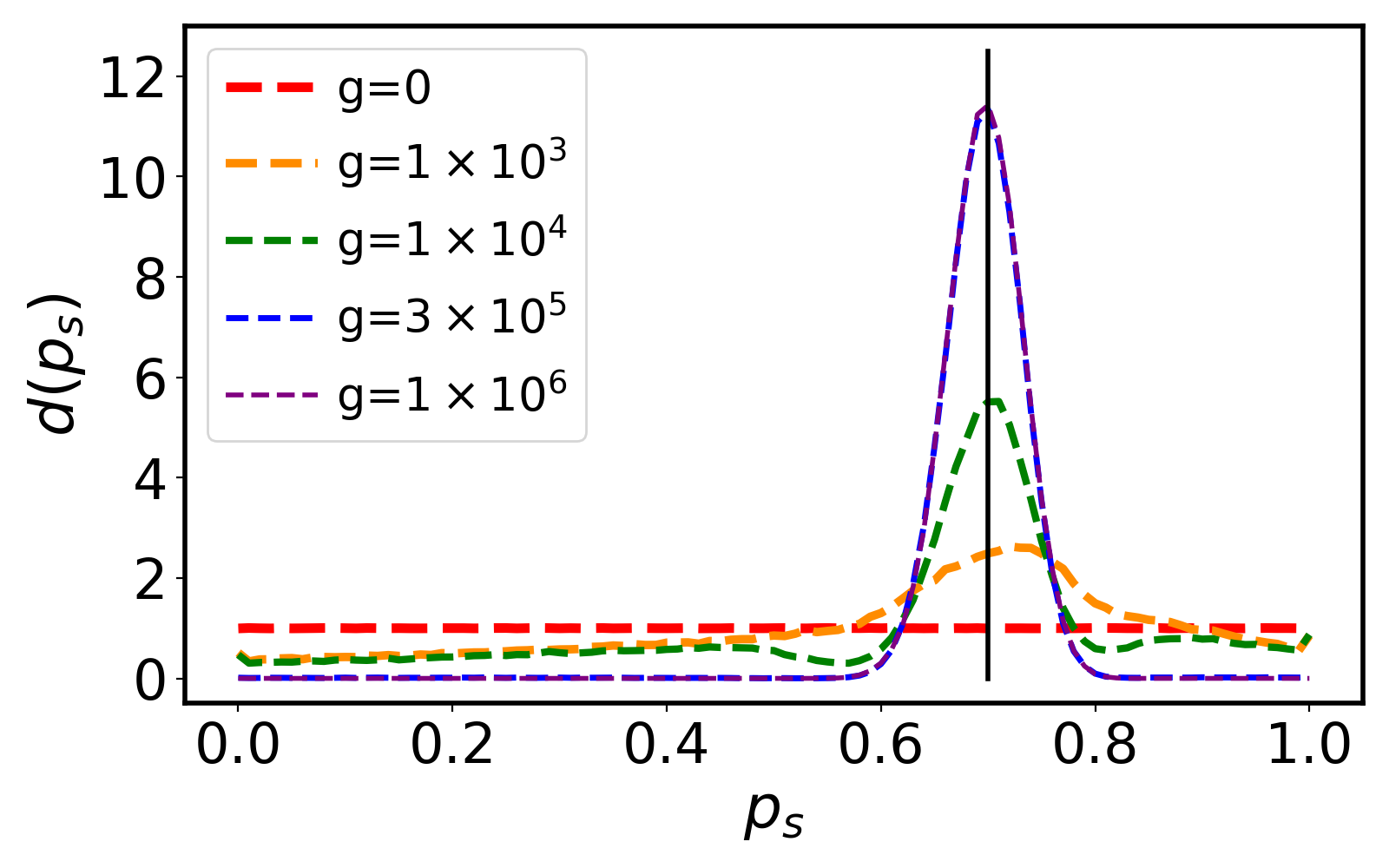}\\
 \caption{
   Distribution of $p_s$ obtained from $M = 10^5$ independent simulations
   with $\ncp=100$ couples,
   starting with an initial uniform distribution over the interval $[0, 1]$,
   for $b_s = 0.7$ and $b_d = 0.3$.
   A peak develops near $\pss = \frac{1-b_d}{2-b_s-b_d} = 0.7$ as generations
   progress. The distribution appears to reach a steady state
   at $g = 3 \times 10^5$, as it remains unchanged thereafter.
   The black vertical line indicates $\pss$ at $p_s = 0.7$.
 }
 \label{f-a1-psdist}
\end{figure}

Figure~A1 shows the distribution of $p_s$ in the population at the generation 
of $g=0$, $g=10^3$, $g=10^4$, $g=3\times 10^5$, and $g=10^6$
for $b_s=0.7$ and $b_s=0.3$ 
form $M=10^5$ different independent simulations. 
We first note that the distribution becomes narrow as
generations progress
and converges to a bell shape distribution centered around
$\pss = \frac{1-b_d}{2-b_s-b_d} = 0.7$.

\begin{figure}[h!]
   \includegraphics[height=120\tm,width=60\tm]{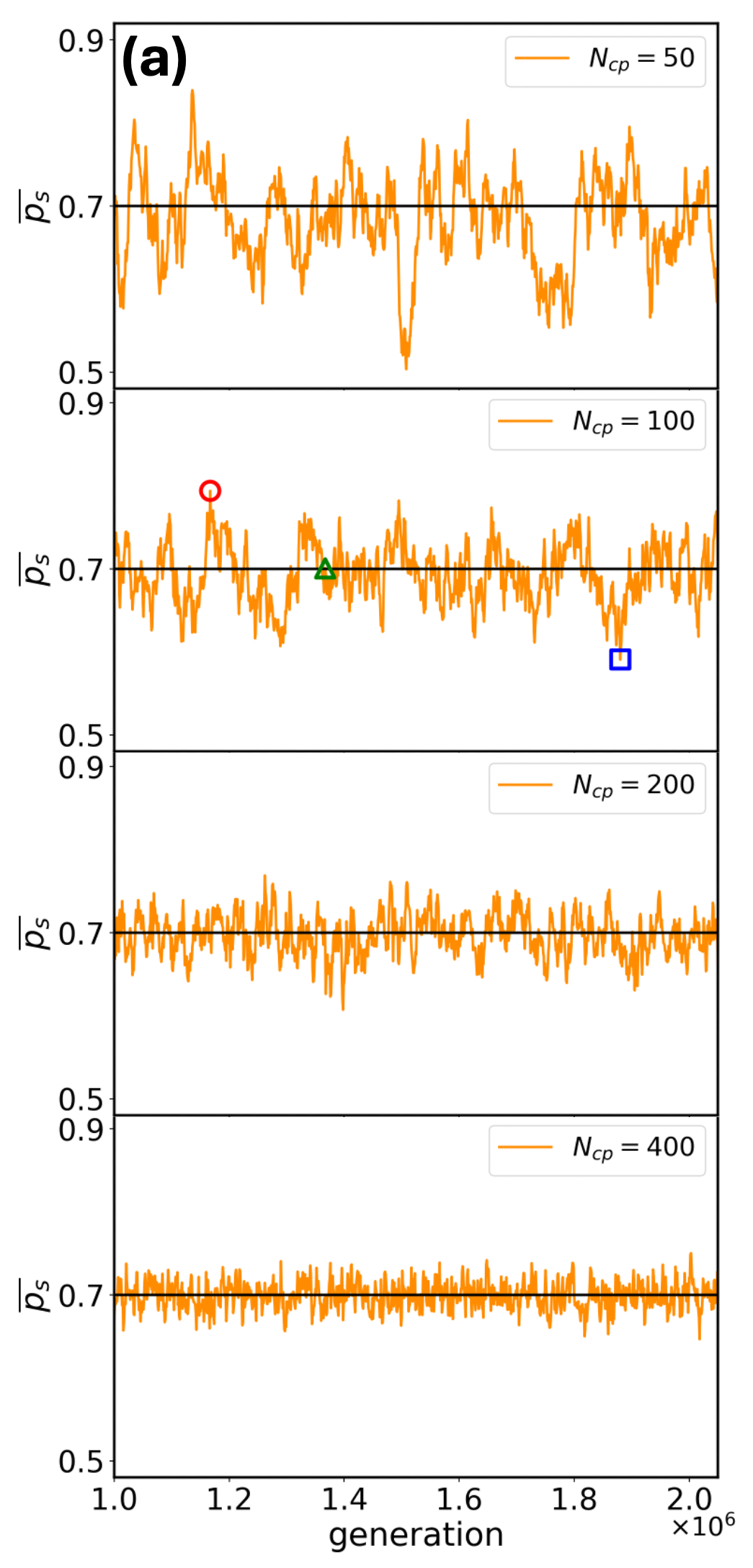}\\
   \includegraphics[height=37.5\tm,width=60\tm]{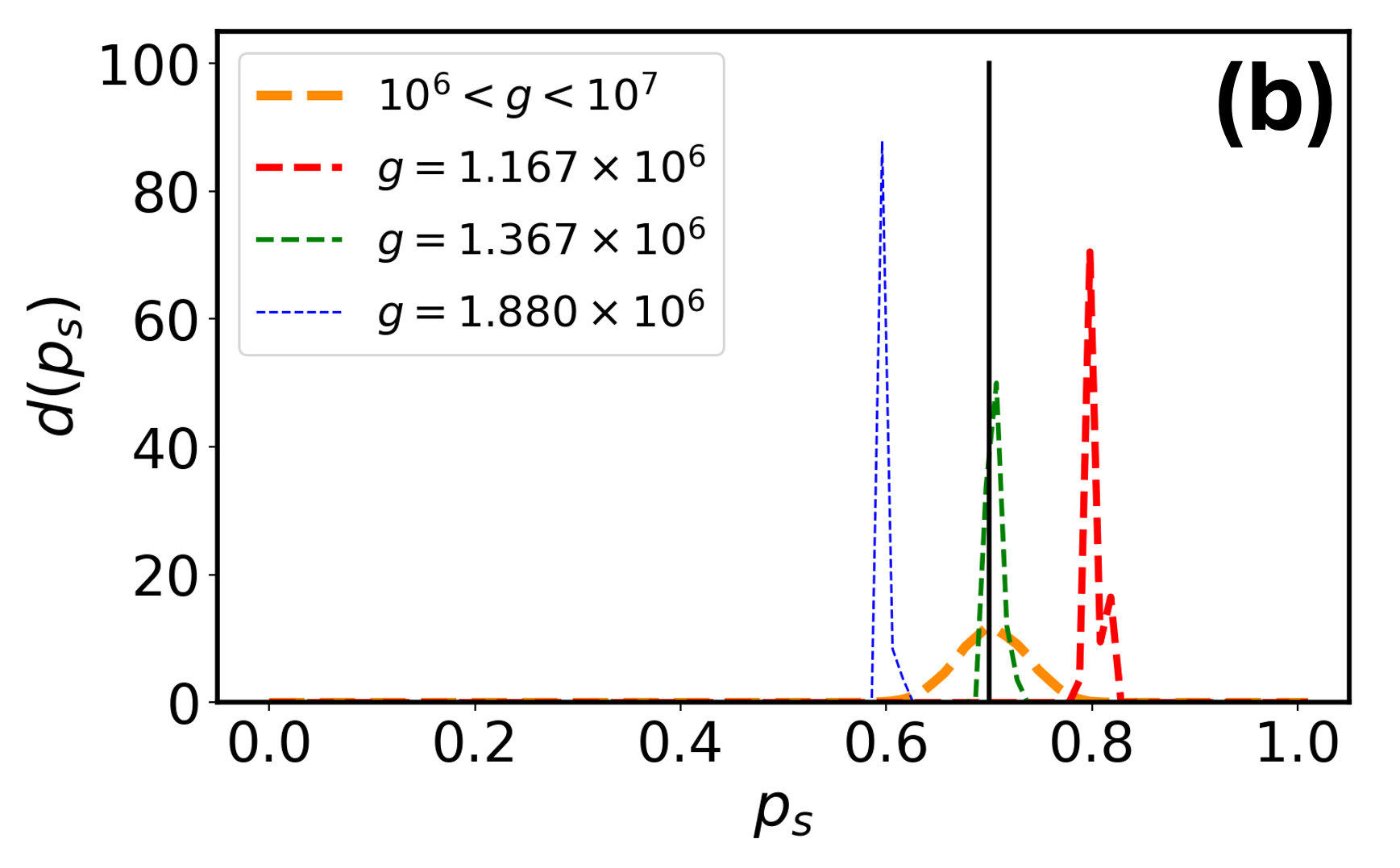}\\
  \caption{
    (a) Time evolution of the average gene value $\psav$ from a single
    simulation with $b_s = 0.7$ and $b_d = 0.3$,
    shown for four populations with different numbers of couples,
    $\ncp = 50$, $\ncp = 100$, $\ncp = 200$, and $\ncp = 400$.
    The values fluctuate around $\pss$ with a relatively long decorrection time.
    (b) The red, green, and blue graphs represent the $p_s$ distributions,
    $d(p_s)$ at three specific points marked on the $\ncp=100$ graph in (a)
    with the same color. 
    The overall $p_s$ distribution from generation $10^6$ to $10^7$ is shown
    as the orange graph.
    The black vertical line represents $\pss$ at $p_s = 0.7$.
  }
  \label{f-a2-time-cor}
\end{figure}

Note that the distribution in Fig.\ref{f-a1-psdist} is obtained from the
son probabilities of individuals across many independent simulations.
In a single simulation, the $p_s$ distribution continues to fluctuate over generations,
even after the overall distribution from multiple simulations has stabilized.
For example, for $g > g_s = 3 \times 10^5$, while the ensemble distribution remains unchanged,
individual simulations still show variations, as illustrated in Fig.~\ref{f-a2-time-cor}.

Figure~\ref{f-a2-time-cor}(a) illustrates the time evolution of the average gene value, $\psav$,
of son probability, defined as,
\beqa
 \psav
 \a= \frac{1}{N} \sum_{i=1}^N p_s,
\label{e-a1-ave}
\eeqa
from a single simulation for four populations with different numbers of couples,
$\ncp = 50$, $\ncp = 100$, $\ncp = 200$, and $\ncp = 400$,
with $b_s = 0.7$ and $b_d = 0.3$. 
The results show that the values fluctuate around $\pss$,
with the magnitude of fluctuations decreasing as the population size increases.
Additionally, these fluctuations are not random from generation to generation,
instead, they exhibit time correlation.
If the value in the current generation is above the average,
it is likely to remain above the average in the next generation.
To quantify this correlation, we calculate the autocorrelation
function over all time lags, $\xi$:
\beqa
 C(\xi)
 \sa= \frac{\frac{1}{G-g_s-\xi}\sum_{g=g_s+1}^{G-\xi}
          \pas{\psav(g) - \psavg}\pas{\psav(g+\xi) - \psavg}}
      {\frac{1}{G-g_s}\sum_{g=g_s+1}^{G} \pas{\psav(g) - \psavg}^2}.
      \nonumber \\
 \label{e-a1-acf}
\eeqa
The decorrelation time, $\tau$, is then defined as the smallest lag $\xi$ for
which $C(\xi) < e^{-1}$.
This measures how long the system retains memory of its previous state.
For the $\psav$ shown in Fig.~\ref{f-a2-time-cor}(a),
we find that $\tau$ is $2.4\times10^4$, $1.2\times10^4$, $0.7\times10^4$, and $0.3\times10^4$ for populations with
$\ncp = 50$, 100,  200, and 400 couples, respectively.
In the subfigure for $\ncp = 100$ in Fig.\ref{f-a2-time-cor}(a),
we mark the $\psav$ values at three specific generations,
which are separated by much more than the decorrelation time,
$g_1 = 1.167\times10^6$, $g_2 = 1.367\times10^6$, and $g_3 = 1.880\times10^6$, by red circle, green triangle,
and blue square, respectively.
The distributions of son probability, $p_s$, at these specific generations
are shown in Fig.\ref{f-a2-time-cor}(b),
along with the overall $p_s$ distribution from generations
$10^6$ to $10^7$ (dashed orange).
At each individual generation, the $p_s$ distribution is narrow,
and its average deviates significantly from $\pss$.
However, when overlapping distributions from many generations
(much longer than the decorrelation time) are combined,
we obtain a Gaussian-like distribution centered at $\pss$.

\begin{figure}[h!]
  \includegraphics[height=37.5\tm,width=60\tm]{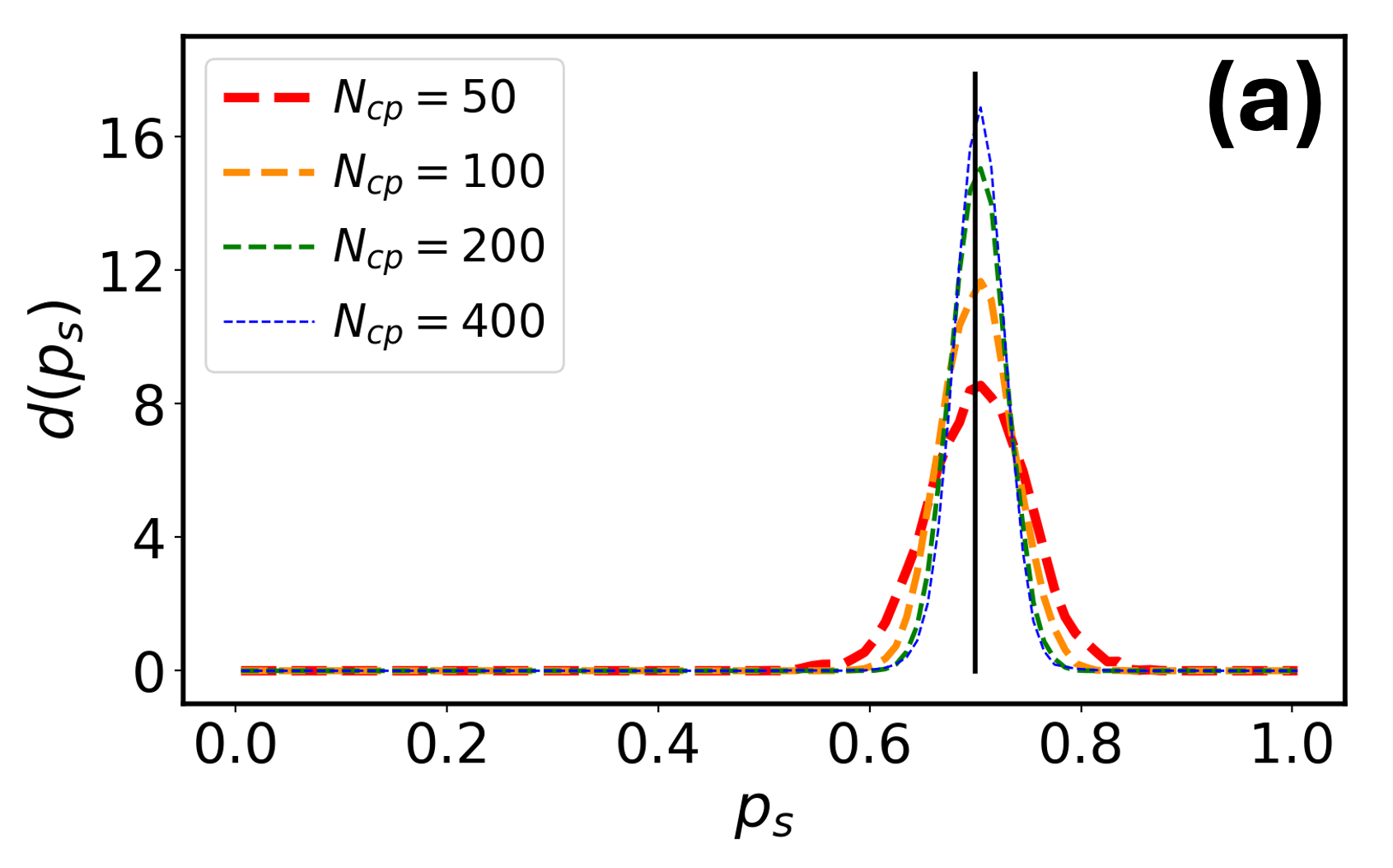}\\
  \includegraphics[height=37.5\tm,width=60\tm]{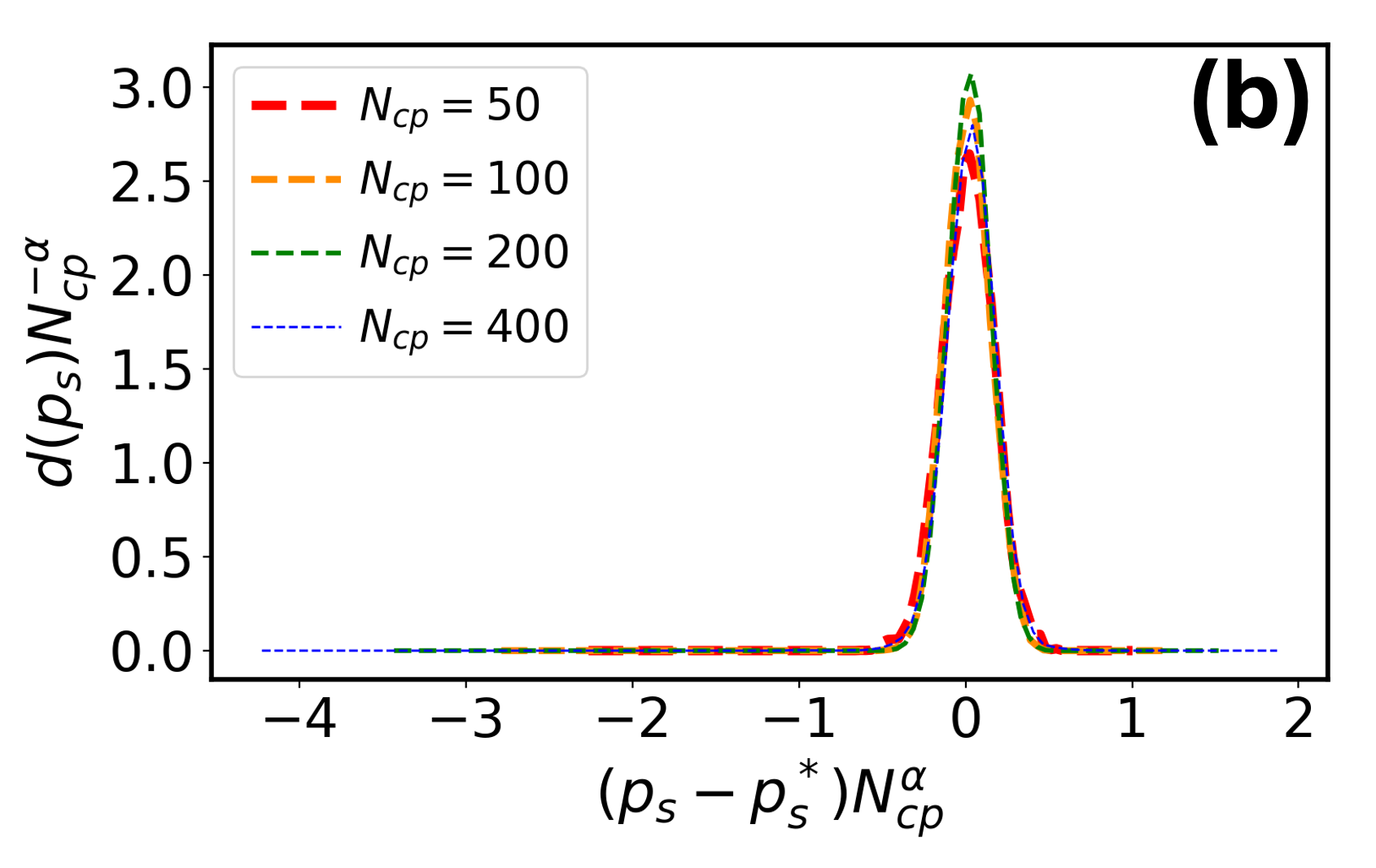}\\
  \caption{
    (a) Distribution of son probability, $p_s$, denoted as $d(p_s)$, from generations
    $10^6$ to $10^7$ for populations with $\ncp = 50$, $\ncp = 100$, $\ncp = 200$, and $\ncp = 400$,
    with $b_s = 0.7$ and $b_d = 0.3$.
    The distributions are Gaussian-like, centered at $\pss = 0.7$, and become narrower with
    increasing $\ncp$.
    (b) Rescaled distributions plotted as $d(p_s) \ncp^{-\alpha}$ against $(p_s - \pss)\ncp^\alpha$ with
    $\alpha=0.3$,
    collapsing onto a single universal curve.
  }
  \label{f-a3-pdf}
\end{figure}

Figure~\ref{f-a3-pdf}(a) presents the overall $p_s$ distribution,
$d(p_s)$ from generations $10^6$ to $10^7$ for four populations with different numbers of couples,
$\ncp = 50$, $\ncp = 100$, $\ncp = 200$, and $\ncp = 400$, with $b_s = 0.7$ and $b_d = 0.3$.
In all cases, the distributions exhibit a Gaussian-like shape centered at $\pss = 0.7$.
As the population size (number of couples) increases, the distribution narrows,
and the peak height correspondingly increases.
When the distributions are rescaled by normalizing the height by $\ncp^{-\alpha}$
and the width by $\ncp^\alpha$, i.e., plotting $d(p_s) \ncp^{-\alpha}$ against $(p_s - \pss)\ncp^\alpha$
they collapse onto a single distribution curve with $\alpha=0.3$,
as shown in Fig.~\ref{f-a3-pdf}(b).
This suggests that as $\ncp$ approaches infinity, the distribution converges to
a Dirac delta-like function.

\section{Appendix B: Proof of $\pss$ as an Evolutionary Stable Strategy}

We define fitness based on the ratio of genes in matched pairs rather than in the entire population.
The son probability $p_s$ is considered an Evolutionarily Stable Strategy (ESS) if no individual with
a son probability $p'_s\not=p_s$ can achieve higher fitness when introduced into a population where all
individuals carry the gene $p_s$.
With this definition, we prove that $\pss = \frac{1-b_d}{2-b_s-b_d}$ is the only ESS for given values of $b_s$ and $b_d$.
Moreover, we show that $\pss$ is not only an ESS but also globally stable
by showing that individuals with $\pss$ always have higher fitness than those with any other son probability,
regardless of their frequency, in a population consisting of only two gene types, $\pss$ and $\pps$.

As in the main text, we consider a case where $b_s$ and $b_d$ are fixed.
Now, suppose a population consists of individuals carrying either gene A
with $p_s$ or gene B with $p_s + \Delta$.
Let $m^A = \frac{M^A}{M^A+M^B}$ and $f^A = \frac{F^A}{F^A+F^B}$
be the proportions of males and females with gene A, respectively.
Following Eq.~(\ref{e-gevo}), the proportion of individuals selected
as parents in the next generation who carry gene B,
denoted as $\cbt$, can be expressed as a function of $p_s$, $\Delta$, $m^A$, and $f^A$.
We determine the value of $\Delta$ that maximizes $\cbt$ for given $p_s$, $m^A$, and $f^A$.
When $\Delta = 0$, genes A and B are identical, so they have the same fitness.
If $\dm \neq 0$, gene B, with $p_s + \dm$, must have a higher fitness than gene A.
Therefore, $p_s$ is not stable at the given $m^A$ and $f^A$ when $\cbt$ reaches
its maximum at $\dm \neq 0$.
Consequently, $p_s$ is not an ESS if $\dm \neq 0$
in the limit of both $m^A$ and $f^A$ goes to 1.
(or for the introuction of small $m^B$ or $f^B$.)

Using Mathematica~\cite{mathematica,lee2025github-math},
we find that $\pd{\cbt}{\Delta} \neq 0$
at $\dm=0$ for all values of $m^A$ and $f^A$, unless $p_s = \pss$.
Moreover, when $p_s = \pss$, we confirm that $\dm = 0$ for all values of $m^A$ and $f^A$.
Hence, $\pss = \frac{1-b_d}{2-b_s-b_d}$ is not only the unique ESS for given values of
$b_s$ and $b_d$ but also the globally stable son probability.

\section*{Acknowledgments}
We thank Eunseo Gwon for her valuable comments on this study. 
This work was supported by the National Research Foundation of Korea(NRF)
grant funded by the Korea government(MSIT) (No. RS-2024-00359230).

\bibliography{00a}


\end{document}